\documentstyle[11pt,newpasp]{article}
\def\mass{{\rm M}}
\def\msol{\mass_\odot}
\def\zform{z_{\rm f}}

\def\b1eff{\overline{b}_1}
\def\b2eff{\overline{b}_2}

\begin{document}

\title{Dark Matter Halos Merging Trees: the Merging Cell
    Model in a CDM Cosmology}

\author{B. Lanzoni\altaffilmark{1}, G. A. Mamon\altaffilmark{1,2},
       B. Guiderdoni\altaffilmark{1}}
\affil{$^1$ Institut d'Astrophysique de Paris, (CNRS UPR 0341), Paris,
       France}
\affil{$^2$ DAEC (CNRS UMR 8631), Observatoire de Paris, Meudon, France}





\section{Contents} 
We have constructed the merging history of dark matter halos in a SCDM 
cosmology, by means of a the {``Merging Cell Model''} (MCM; ref: RT96).
It is based on the linear theory of growth of density fluctuations, and the
Top-Hat model, and it takes into account mergers between halos through
simplified criteria. 

The halo mass function, the progenitors and children mass distribution, 
the distribution of formation times, and the halo autocorrelation 
function have been computed. 
To test the reliability of the model, results have been compared to the
analytical predictions of the extended Press \& Schechter (ePS) theory, that
in turn well describes results of N-body simulations.

A general good agreement has been found, thus providing us with a reliable 
tool for rapid
simulations of galaxy formation and evolution within hierarchical scenarios. 
The MCM is particularly suitable for studying galaxy clusters at low
redshift, and the population of Lyman-break galaxies at high z.



\section {Some results}
Some results of the comparison between MCM and analytic
predictions are shown in the figures below, while an extensive discussion is
in Lanzoni et al. (1999). 
A SCDM cosmology is adopted, with the following values of the cosmological
parameters: 
H$_0 = 100\,h\,{\rm km\,\, s}^{-1} {\rm
Mpc}^{-1}$, $h=0.5$, $\Omega_0=1$, $\Omega_{\rm b}=0.05$,
$\Omega_{\rm\Lambda}=0$, $\sigma_{8/h}=0.67$.
The MCM results are averaged over 10 different realisations with $256^3$ base
cells, in a comoving volume of $50\,h^{-1}$ Mpc side.
The mass resolution is of about $2\!\times\!10^{9}\,h^{-1}\msol$, and the
most massive halo typically has a mass of about
$5\!\times\!10^{14}\,h^{-1}\msol$.  

{\bf Fig.1:} Differential mass function of halos
at $z=0, 3$. 
Results from the MCM model (histograms) are compared to the analytic
prediction of the ePS theory (e.g., LC94; dotted
lines), and to the LS98 mass function that better
describes results from N--Body simulations (solid lines). 
An overall good agreement is found, even if a lack of low-mass halos in the
MCM model is apparent. Also too few high-mass halos are found, especially at
high $z$. 

{\bf Fig.2:} Differential probability distribution of formation redshifts
$\zform$, for halos in two mass ranges at $z=0$.
Solid curves are the analytic predictions (LC94).
Within the error bars, a very good agreement is found for the most
massive objects, while severe discrepancies are found for small objects.

{\bf Fig.3:} 
Autocorrelation function of halos in two mass ranges, 
selected at $z=3$. 
MCM results {circles} are compared to the Jing's fitting formula (Jing 1998;
solid lines), that provides good fits to N-body simulations. 
A very good agreement is apparent.

\begin{figure}
\includegraphics{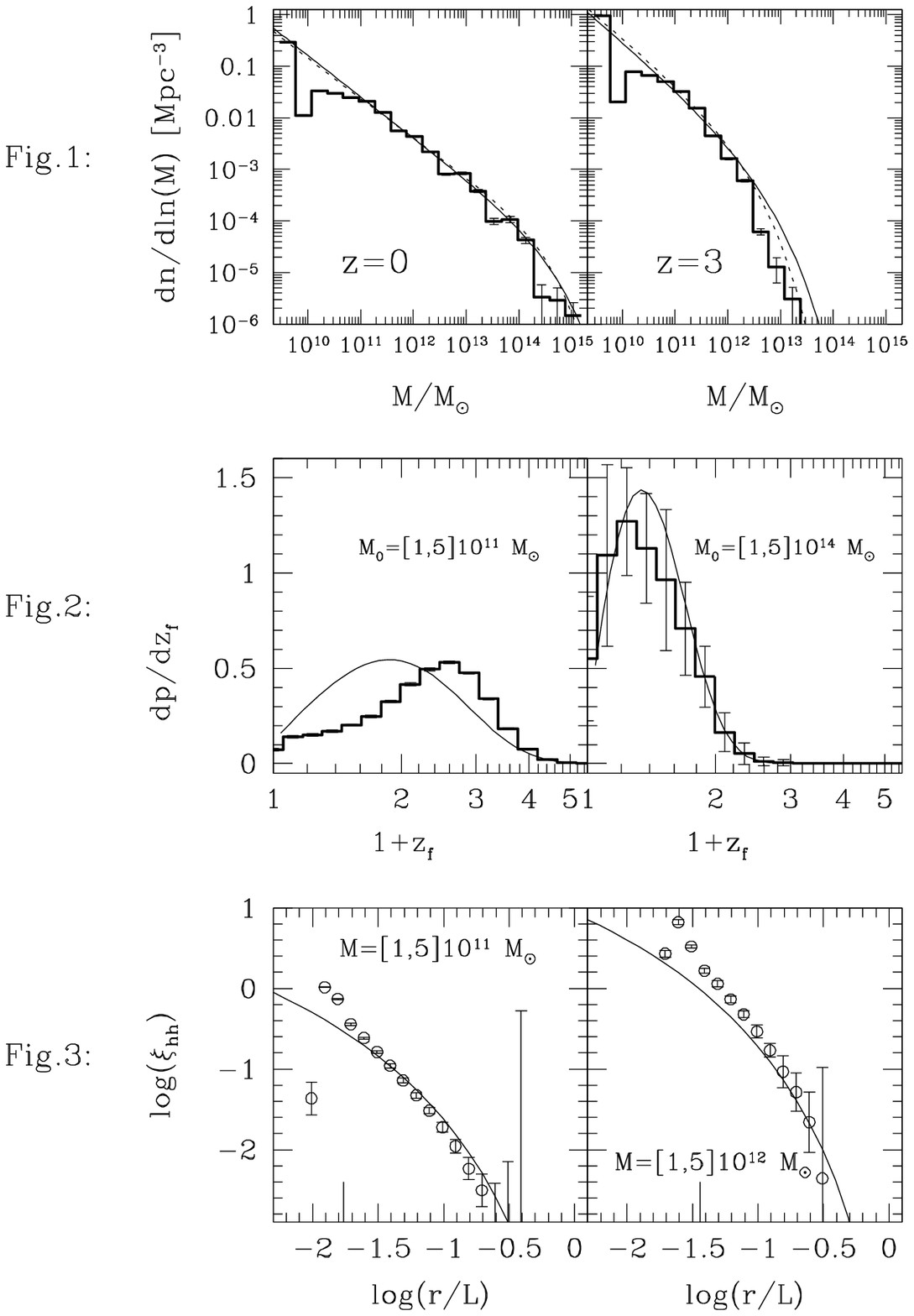}
\end{figure}

\end{document}